\documentclass[prd,preprint,eqsecnum,nofootinbib,amsmath,amssymb,
               tightenlines]{revtex4}

\usepackage{graphicx}% Include figure files
\usepackage{bm}% bold math
\def\Tr{\: {\rm Tr} \:}

\def\x{{\bf x}}
\def\p{{\bf p}}
\def\ps{{\bf ps}}

\def\F{{\mathcal F}}
\def\N{{\mathcal N}}

\def\Im{{\rm Im}}

\def\st{\begin{equation}}
\def\stp{\end{equation}}
\def\bg{\begin{eqnarray}}
\def\nd{\end{eqnarray}}
\def\Eq#1{Eq.~(\ref{#1})}
\def\app#1{Appendix~\ref{#1}}
\def\Fig#1{Fig.~\ref{#1}}
\def\Sect#1{Section~\ref{#1}}
\def\Ref#1{Ref.~\cite{#1}}

\def\llangle{\left\langle}
\def\rrangle{\right\rangle}

\def\tr{\operatorname{tr}}

\def\N{\mathcal{N}}
\def\w{\mathfrak{w}}

\def\tr{\mbox{tr}}
\def\gsim{\mbox{~{\protect\raisebox{0.4ex}{$>$}}\hspace{-1.1em}
	{\protect\raisebox{-0.6ex}{$\sim$}}~}}

\def\nott#1{\setbox0=\hbox{$#1$}                % set a box for #1 
   \dimen0=\wd0                                 % and get its size
   \setbox1=\hbox{/} \dimen1=\wd1               % get size of /
   \ifdim\dimen0>\dimen1                        % #1 is bigger
      \rlap{\hbox to \dimen0{\hfil/\hfil}}      % so center / in box
      #1                                        % and print #1
   \else                                        % / is bigger
      \rlap{\hbox to \dimen1{\hfil$#1$\hfil}}   % so center #1
      /                                         % and print /
   \fi}                                         %

\advance\parskip 1.9pt
\advance\voffset -0.2in

\def\st{\begin{equation}}
\def\stp{\end{equation}}
\def\bg{\begin{eqnarray}}
\def\nd{\end{eqnarray}}

\begin{document}

\vspace*{1cm}

\title{Heavy Quark Diffusion in Strongly Coupled $\N=4$ Yang Mills}
%% Or Something Like That

\author{Jorge Casalderrey-Solana}
\author{Derek Teaney}
\affiliation
    {%
    Department of Physics \& Astronomy,
    SUNY at Stony Brook,
    Stony Brook, New York 11764, USA
    }%
\date{\today}

%%\pacs{Valid PACS appear here}
%% God forbid we need either PACS numbers or keywords.

%%\keywords{Suggested keywords}%Use showkeys class option if keyword
%                              %display desired

\begin{abstract}
We express the heavy quark diffusion coefficient 
as the temporal variation of a Wilson line along the 
Schwinger-Keldysh contour. This generalizes
the classical formula for diffusion as a force-force correlator 
to a non-abelian theory. 
We use this formula to compute the diffusion 
coefficient  in strongly coupled $\N=4$
Yang-Mills by studying the fluctuations of a string in 
$AdS_5\times S_5$. The string solution spans the 
full Kruskal plane  and gives access to contour correlations.
The diffusion coefficient is  
$D=2/\sqrt{\lambda} \pi T$ and is therefore parametrically
smaller than momentum diffusion, $\eta/(e+p)=1/4\pi T$. 
The quark mass must be much greater than $T\sqrt{\lambda}$
in order to treat the quark as a heavy quasi-particle.
The result is discussed in the context of the RHIC experiments.
%We also find the string in $AdS_5 \times S_5$
%which corresponds to a heavy quark moving with velocity $v$ 
%in the $\N=4$  plasma. This is a first step to understanding
%the dynamics of a heavy quark which is not at rest in the
%medium.  
\end{abstract}
\maketitle
%%%%%%%%%%%%%%%%%%%%%%%%%%%%%%%%%%%%%%%%%%%%%%%%%%%%%%%%%%%%%%%%%%%%%%%%%%%%%%%%

\section{Introduction}

The experimental relativistic heavy ion program has produced a variety
of evidences which suggest that a Quark Gluon Plasma (QGP) has been
formed at the Relativistic Heavy Ion Collider (RHIC)
\cite{Bellwied:2005kq,Adcox:2004mh}. One of the most exciting
observables is the medium modifications of heavy quarks. 
In particular the electron spectrum from the semi-leptonic
decays of heavy quarks is substantially
suppressed relative to scaled proton-proton collisions \cite{Adler:2005xv,Bielcik:2005wu}.
Furthermore preliminary measurements indicate that the heavy quark
elliptic flow is significant although less than the light 
hadron elliptic flow \cite{Akiba:2005bs}. 

A variety of phenomenological
models have estimated how the transport mean free path of heavy
quarks in the medium is ultimately reflected in the
suppression factor and elliptic flow \cite{Moore:2004tg,Molnar:2004ph,Zhang:2005ni}. 
The result of these model studies is best expressed in terms of the 
heavy quark diffusion coefficient. 
(In a relaxation time approximation the diffusion coefficient is related
to the equilibration time, $ \tau_R^{\rm heavy}=\frac{M}{T}D$.)
There is a sense from the models that if the diffusion coefficient of the heavy
quark is greater than 
\[
D \gsim \frac{1}{T} \,,
\]  
the heavy quark medium modifications
will be small and probably in contradiction with current data. 
This  interpretation of the RHIC results is perhaps 
too naive since the diffusion coefficient dictates the dynamics
of  non-relativistic heavy quarks. Diffusion may  be irrelevant
for the dynamics of the mildly relativistic heavy quarks measured at RHIC where  radiative energy loss may be significant 
\cite{Djordjevic:2005db,Armesto:2005zy}. Nevertheless, the diffusion 
coefficient is a fundamental parameter of the plasma and 
is essential to any discussion of the RHIC heavy flavor data.

Unfortunately, for 
the experimentally relevant range of energy densities
the QGP is not weakly coupled, and 
it is not easy to determine this transport coefficient. 
Ideally, the diffusion 
coefficient should be measured on the lattice, but this is difficult
 \cite{Petreczky:2005nh,Aarts:2002vx} .
Most theoretical works either compute the diffusion 
coefficient in perturbation theory and subsequently 
extrapolate to strong coupling (see e.g. \cite{Moore:2004tg}), or develop  models for the 
strongly interacting QGP \cite{vanHees:2004gq}. 
In this work we will compute the diffusion coefficient in 
strongly coupled $\N=4$ Super Yang Mills (SYM) where rigorous 
computations are possible if the $AdS/CFT$ conjecture 
is accepted.  
%We will also find the string configuration
%in the $AdS_5 \times S_5$ background
%which represents a heavy quark moving with velocity $v$.
%This configuration will be useful in subsequent studies 
%of the dynamics of  relativistic heavy quarks.
Although this theory is not QCD, computations
in this theory serve as a foil to the extrapolations 
based on weak coupling. 

This conjecture states that for a large number of 
colors, strongly coupled $\N=4$ 
SYM is dual to classical type IIB supergravity on 
an $AdS_5 \times S_5$ background
\cite{Maldacena:1997re,Gubser:1998bc,Witten:1998qj}. 
(For reviews and lectures see \Ref{Aharony:1999ti,D'Hoker:2002aw}.)
The physics of heavy quarks has been studied using 
semi-classical strings \cite{Maldacena:1998im,Drukker:1999zq,Brandhuber:1998bs,Brandhuber:1998er,Rey:1998bq,Rey:1998ik}.
On the gauge theory side a heavy charge 
is  realized  
by breaking the $U(N)$ gauge theory to $U(N-1)\times U(1)$
through the Higgs Mechanism. The resulting $W$ boson transforms
in the fundamental and is heavy if the scalar expectation 
value is large.  
On the gravity side this corresponds to placing
one of the $D3$ branes far from the remaining $N-1$ branes.
The dynamics of the heavy quark 
is dictated by the classical dynamics of the Nambu-Goto string 
stretching in the $AdS_5 \times S_5$ background. 
The first thing computed was the expectation value of
the Wilson loop to find the heavy quark potential \cite{Maldacena:1998im}. The result was immediately extended to
finite temperature \cite{Brandhuber:1998bs,Brandhuber:1998er,Rey:1998bq,Rey:1998ik}. Since then there have been numerous 
studies of the other properties of Wilson Loops both
at zero and non-zero temperature \cite{Aharony:1999ti}.

One of the most interesting transport properties of the $\N=4$ plasma
is the heavy quark diffusion coefficient. Various transport  properties
have been computed using the correspondence leading to the remarkable
conjecture that the shear viscosity to entropy ratio is bounded from
below by $1/4\pi$ \cite{Kovtun:2004de,Buchel:2004qq,Policastro:2001yc,Policastro:2002se}.  This bound is influencing the
interpretation of heavy ion results.  Generally these calculations of
transport in $\N=4$ use the supergravity approximation.
In contrast the computation of heavy quark diffusion
will utilize the
classical string theory directly.  The diffusion computation in $\N=4$ SYM is
therefore  experimentally and theoretically significant.
For other applications of the $AdS/CFT$ correspondence
motivated by RHIC physics see 
Refs.~\cite{Sin:2004yx, Shuryak:2005ia}.
 
The computation proceeds as follows. In \Sect{basic} we 
review  the Langevin  dynamics of the heavy quark. 
In \Sect{kappa_sect}  we show that the 
 strength of the noise in the Langevin process 
 is determined by  real time electric 
field correlators once the heavy quark has been 
integrated out. 
These electric field correlators 
can be expressed as fluctuations of a Wilson line running along the
Schwinger-Keldysh contour.  
These results generalize the classical formula
for the diffusion  as a force-force correlator to
 non-abelian theories.
In \Sect{fluct} we compute 
the fluctuations of the string configuration which corresponds to
this Wilson line running along the contour;  the
relevant string  spans the  Kruskal plane.
%In \Sect{bended_string} 
%we determine the string configuration which corresponds to 
%heavy quark moving with velocity $v$ in the medium. This
%string configuration will be useful for subsequent investigations
%of the transport properties of heavy quarks. 
Finally in \Sect{discuss},
we summarize our results and discuss the implications for
the RHIC experiments.
 
\section{Langevin Dynamics}
\label{basic}

In this section we will discuss the predictions of the 
Langevin process for determining the  retarded correlator. 
We will assume that the Langevin equations 
provide a good 
macroscopic description of the thermalization of heavy particles
\begin{eqnarray}
\label{newton_langevin}
\frac{dx_i}{dt} &=& \frac{p_i}{M} \, , \nonumber \\
\frac{dp_i}{dt} &=& \xi_i(t) - \eta_D p_i \; , \qquad \langle \xi_i(t) \xi_j(t') \rangle = \kappa \delta_{ij}
	\delta(t-t') \, .
\end{eqnarray}
%(For a brief review of the Langevin equation see \Ref{Moore:2004tg})
The drag and fluctuation coefficients are related  by the 
Einstein relation 
\begin{equation}
\label{etad}
\eta_D = \frac{\kappa}{2 MT} \; .
\end{equation}
The drag coefficient $\eta_D$ can be related in turn to the diffusion coefficient
\begin{equation}
\label{DfromKappa}
   D = \frac{T}{M \eta_D} = \frac{2 T^2}{\kappa}  \; .
\end{equation}
For a brief review of the Langevin equation and a derivation 
of these results, see \Ref{Moore:2004tg}.

The relevant time scale for medium correlations in a relativistic plasma is
$\sim D$ which is short compared to the relaxation time of the heavy
particle $(M/T)\, D$.  Furthermore, over the time scale of medium
correlations the quark moves a  negligible distance, $\sqrt{T/M}\,D$.
Thus, for the purposes of calibrating the noise ($\kappa$), 
the mass may be taken to infinity and the heavy quark may be considered
fixed.

More generally, the  microscopic equations of motion 
for a heavy particle in the medium are
\begin{eqnarray}
   \frac{dx^i}{dt} = \frac{p^i}{M} \, ,\\
   \frac{dp^i}{dt} = \mathcal{F}^i(t) \; .
\end{eqnarray}
Compare the response of the Langevin process 
to  the microscopic theory. 
Over a time which is long compared to medium correlations $D$, but 
short compared to the time scale of equilibration $(M/T) D$,  the 
drag term may be dropped, and we find
\begin{equation}
   \int dt \,\int dt' \, \llangle \xi_i(t) \xi_j(t') \rrangle = \mbox{(time)}  \times  \kappa \, \delta_{ij} = 
    \int dt \, \int dt' \,\llangle \mathcal F_i(t) \mathcal F_j(t') \rrangle   \; .
\end{equation}
Taking the force in the $y$-direction for instance, we have
\begin{eqnarray}
\label{ff}
    \kappa & = &  \int dt \, \llangle {\mathcal F}_y(t) {\mathcal F}_y(0)  \rrangle \; .
\end{eqnarray}
We will drop the ``y" in what follows.
In $QED$ this analysis relates the Langevin noise to an
an electric field 
correlator \cite{Forster,BooneYip}.
The next section generalizes
this result to  gauge theories. 

\section{Momentum Diffusion in Gauge Theories }
\label{kappa_sect}

We study the propagation of a heavy quark in a thermal
bath of SU(N) gauge fields.
After a  Foldy-Wouthuysen transformation that
reduces the heavy quark field to a two dimensional spinor field $Q$ \cite{Brown:1979ya}, the
heavy quark effective Lagrangian is
\st
\label{lagrangian}
\mathcal{L}=Q^{\dagger}\left( i\partial_t -M-A_0\right)Q \, .
\stp
From this Lagrangian it is natural to identify the force
 in the Langevin equation with the operator
\st
 \label{operator}
 \mathcal{F} \equiv \int d^3 x \,
                        Q^{\dagger}(t,{\bf x})T^aQ(t,{\bf x})\,{ E}_a(t,x) \, ,
\stp
where $E$ is the chromo-electric field in the $\hat{y}$ direction.
 The momentum diffusion coefficient $\kappa$ is given by  \Eq{ff}
\begin{eqnarray}
\kappa &=&\int dt  
 \llangle
\mathcal{F}(t)
\,
\mathcal{F}(0)
\rrangle_{HQ} \, ,
\end{eqnarray}
where the average should be understood as a thermal
average in the presence of a heavy quark -- the meaning 
of this average is clarified below and in \app{partition}.
%Note that we have divided by the total number of quarks in the 
%bath
%\st
%N_{\scriptstyle{\rm{Q}}}=\int d^3 x \llangle
%                                    Q^{\dagger}_i(t,{\bf x})Q_i(t,{\bf x}) 
%                                    \rrangle
%\stp
%We are therefore computing the thermal corelation functions
%the operator
%\begin{eqnarray}
%\label{operator}
%{\mathcal F}(t)&=& \int
%d^3\x Q^{\dagger}(t,{\bf x}){\bf E}(t,{\bf x})Q(t,{\bf x})
%\end{eqnarray}

\begin{figure}
\begin{center}
\includegraphics[width=10cm]{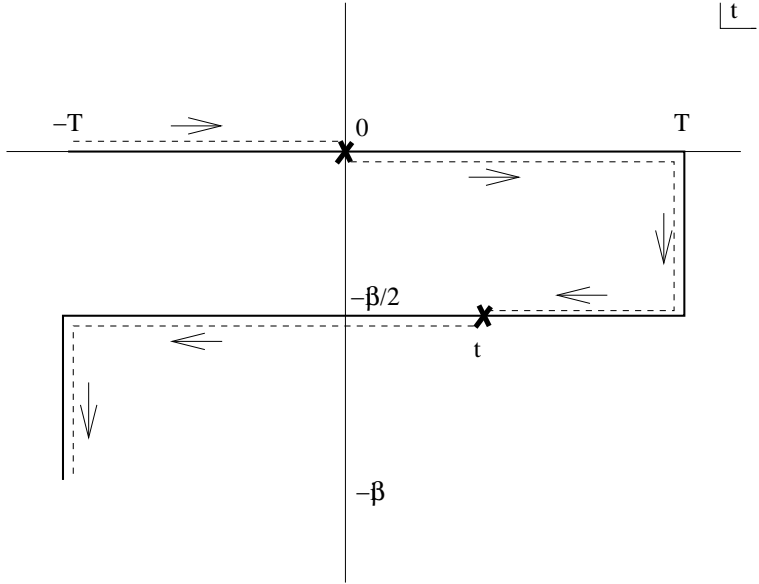}
\caption
{
The Schwinger-Keldysh contour (with $T\rightarrow \infty$). Fields evaluated 
along the real axes are labeled as type 1, while fields evaluated
on the $-i\beta/2$ axis are labeled as type 2.
The crosses indicate the insertion points of the color 
singlet force operator 
$\mathcal{F}$ (Eq. \ref{operator}).  After integrating
out the heavy quark, the crosses indicate insertions
of the electric field, and the dotted lines indicate the 
path of the corresponding links. 
 The electric field insertion 
may be rewritten as a variation (at times $0$ and $t$)
of the Wilson line running along the whole contour. 
}
\label{schwinger}
\end{center}
\end{figure}

In the real time formalism of thermal field theory 
(see e.g. \cite{LeBellac,Gelis:1994dp}),
the quantization of the fields is performed by 
extending the time coordinates into the complex plane along 
the Schwinger-Keldysh contour -- see Fig. \ref{schwinger}.
There are two types of  quantum fields: type 1 fields, evaluated on
the real time axis and type 2 fields, which are evaluated on the second 
$-i\beta/2$ time axis. We therefore define the correlation functions
\begin{eqnarray}
iG_{11}(t,t') &=& \llangle\, T\, \F_1(t)\, \F_1(t')\,\rrangle_{HQ} \, ,    \\
iG_{12}(t,t') &=& \llangle\, \F_2(t')\, \F_1(t) \,\rrangle_{HQ} \, , \\
iG_{21}(t,t') &=& \llangle\, \F_2(t)\, \F_1(t') \,\rrangle_{HQ} \, , \\
iG_{22}(t,t') &=& \llangle\, \tilde{T}\, \F_2(t)\, \F_2(t') \, \rrangle_{HQ} \, ,
\end{eqnarray}
where $T$ and $\tilde{T}$ denote time and anti-time ordering  respectively, and
\begin{eqnarray}
     \F_1(t) &=& e^{+iH t } \F(0) e^{-iH t} \, , \\
     \F_2(t) &=& e^{+iH\,(t - i\beta/2)} \F(0) e^{-i H\,(t - i\beta/2)} \, .
\end{eqnarray}

Temporal correlators of the operator $\F$
in the presence of an external heavy quark
verify the Kubo-Martin-Schwinger 
(KMS) relations,  provided single particle 
states may be considered complete, $i.e.$  gauge fields 
are not so strong as to create heavy quark-antiquark pairs  
and the density of heavy quarks is small -- see \app{partition}. 
In particular we define the Retarded Green function
\st
i G_R(t)=\theta(t) \llangle[\mathcal{F}(t),\mathcal{F}(0)]\rrangle_{HQ} ,
\stp
and relate the  remaining Green functions by inserting 
complete sets of states 
\bg
\label{G11}
 i G_{11}(\omega) &=&  i \mbox{Re}\,  G_{R}(\omega) -  \coth\left(\frac{\omega}{2T} \right) \, \mbox{Im} G_{R}(\omega)  \, ,\\
\label{G12}
 i G_{12}(\omega) &=&  i G_{21}(\omega) =  
 - \frac{2 e^{\frac{-\beta \omega}{2} } } {1 - e^{-\beta \omega} } \, \mbox{Im} G_{R}(\omega)  \, ,\\
\label{G22}
 i G_{22}(\omega) &=& -i \mbox{Re}\, G_{R}(\omega) - \coth\left(\frac{\omega}{2T} \right)\,\Im G_{R}(\omega) \, .
\nd
Using the fluctuation dissipation theorem, $\kappa$ is related to 
zero frequency limit of the retarded correlator
\st
\kappa= 
 \lim_{\omega \rightarrow 0} -\frac{2 T}{ \omega} \Im G_R(\omega) \, . 
\stp
This  can  be determined from $iG_{12}(\omega)$  for instance.

For a Yang-Mills theory with  external heavy quark the partition function is 
\cite{McLerran:1981pb}
\bg
\label{ZHQPI}
Z_{HQ} &=&\int d^3 x
\llangle 
\frac{1}{\det\left[i\partial_{t_{C}}- M- A_0 - i\epsilon  \right]} \right.   \nonumber \\
 & &\left. \times \int [D\,Q][D\,Q^{\dagger}] e^{
                      i\int_C Q^{\dagger}\left(i\partial_{t_C}-M-A_0 -i\epsilon \right)Q}\, Q_i(\x, -i\beta) Q_i^{\dag}(\x, -T)\rrangle_{YM}
\, ,
\nd
where the  integration is performed along the Schwinger-Keldysh contour. 

The heavy fermion may be integrated out of this expression
of the partition function.
The propagator of the heavy quark in a fixed gauge 
background is computed from the Green's function
\st
\label{GF}
\left(i\frac{\partial}{\partial t_C}-M-A_0 - i\epsilon \right) G\left({\bf x},t_C ;{\bf x'},t_C'\right)
=i\delta\left({\bf x}-{\bf x'}\right)\delta\left(t_C-t_C'\right)
\, .
\stp
With this propagator,
the partition function of the heavy quark yields the 
Polyakov loop along the contour
\footnote
{
$V_{\ps}$ is the phase space volume
\st
V_{\ps}=V\int \frac{d^3\, p}{(2\pi)^3}=V\,\delta^3({\bf 0}) \, .
\stp
} 
\st
\label{ZHQ}
    Z_{HQ}= V_{\ps} e^{-\beta M} \llangle W_{C}[0] \rrangle \, .
\stp
Usually the contour of the heavy quark partition function 
is taken straight down the imaginary 
axis to $-i\beta$, yielding the usual Polyakov loop. 
However this is unnecessary  and the answers are the same.

Similarly, the heavy fermion fields may be integrated 
out of contour ordered force-force correlators
\st
\label{diffWL}
\llangle T_C[\mathcal{F}(t_C)\mathcal{F}(0)]
\rrangle_{HQ}= \frac{1}{\llangle W_C[0] \rrangle } 
\llangle \tr{\left[U(-T -i\beta,t_C)\,  E(t_C) \,
  U(t_C,0)\, E (0)\,  U(0, -T)\right]}\rrangle \; ,
\stp
where the tranporters $U(t_f,t_i)$ are calculated along the 
Schwinger-Keldysh contour with $t_i$ and $t_f$ the initial and final contour times. The denominator ${\llangle W_C[0]\rrangle}$ stems
from the partition function \Eq{ZHQ}.
%As the fields are 
%periodic in $i\beta$,  the path is  closed  as 
%shown in Fig. \ref{schwinger}. 
%The prefactor  $exp{\{-M\beta\}}$
%corresponds to number of Heavy Quarks on the bath for a free theory. 
%However, in the interacting theory de total number of quarks is given
%by
%\st
%V_{\scriptstyle{\rm{ps}}}e^{-\beta M} \llangle W[0]\rrangle
%\stp
%w?here $W[0]$ is the Wilson loop along the Scwhinger-Kedlish contour 
%(?wich coincides with the Polyakov Loop). We will 
%d?ivide the final expression by this factor.
This correlator of electric fields can be generated  by starting  with a
Wilson loop  running along the contour and inserting 
small notches at contour times $t=0$ and  $t=t_C$ as
shown in \Fig{schwinger}.  

Generalizing this procedure, we first introduce a source $\delta y(t_C)$ for the force 
operator 
\begin{eqnarray}
\frac{1}{Z_{HQ} [0]} Z_{HQ}[\delta y]=
\llangle T_C 
                     e^{i\int_C dt\, \delta y(t)\, \mathcal{F}(t)}
                      \rrangle_{HQ}
\, .
\end{eqnarray}
%We now show this is equal to a variation 
%of a Wilson line
%\[
%\mathcal{Z}[\delta y]= \llangle T_C 
%                     e^{i\int_C dt \delta y(t_C) \int d^3x \mathcal{F}(t_C,x)}
%                      \rrangle_{YM+HQ} = \frac{1}{\llangle W[0] \rrangle} \llangle W[\delta y] \rrangle  
%\]
This is accounted for by 
modifying  the heavy quark Lagrangian 
\st
\mathcal{L}_{HQ}=Q^{\dagger}\left(
                              i\partial_{t_C}-M-A_0-\delta y(t_C)\,E(t_C)
                              \right)
                 Q
\, .
\stp
As in \Eq{diffWL}, we integrate the partition function in the 
presence of a heavy quark
\st
Z_{HQ}[\delta y]=  V_\ps e^{-\beta M} \llangle W_{C}[\delta y]\rrangle
\, ,
\stp
where 
\st
W_{C}[\delta y]=T_C \exp{\left\{-i\int_{C} dt 
                               \left(A_0 + \delta y(t) E(t)\right)
                                   \right\}}
\, .
\stp
This object can be easily expressed as
\footnote
{
$\delta y$ is assumed to be small.
} 
\st
\label{AlmostWilson}
W_{C}[\delta y]=
T_C \exp{\left\{-i\int_C dt\, 
                               A_0 
                                   \right\} }
T_C \exp{\left\{i\int_C dt\, 
                               \delta y(t) \tilde{E}(t)
                                   \right\} } \, ,
\stp
where $\tilde{E}$ is the dressed field strength
\st
\label{dressedF}
\tilde{E}(t_C)=U(-T, t_C) \,E(t_C)\, U(t_C, -T) \, ,
\stp
$i.e.$ the transporter starts at an initial time $-T$, runs along the Schwinger-Keldysh 
contour from the initial point up to the contour time $t_C$,
and returns to the initial time,  $-T$. 
By means of the non Abelian Stokes theorem \cite{Bralic:1980ra} or some thought,  
it can be shown that
\Eq{AlmostWilson} is the Wilson line
along the time contour with a deformation in the $\hat{y}$ direction
given  by $\delta y(t_C)$
\st
W_C[\delta y] = T_C \exp{\left\{-i\int_C dt \left(
                               A_0 + \delta \dot{y} A_y
                                    \right)
                                   \right\}} \, .
\stp
In summary, the path $\delta y(t)$ of a Wilson loop
running around the Schwinger-Keldysh contour is the source
for contour ordered force operators, $\mathcal F$. 
In real time thermal field theory 
it is custumary to break up the source into $1$ type sources 
and $2$ type sources with the understanding that variations
of the vertical part are not considered.  The source
$\delta y_1(t)$ and $\delta y_2(t)$ are variations of the 
the Wilson loop on  the $1$ and $2$ axis respectively 
\st
     \llangle T e^{i\int dt 
\, \delta y_1(t) \, {\mathcal F}_1(t)}\; \tilde{T} e^{-i\int dt' \, \delta y_2(t') \,{\mathcal F}_2(t')  }  \rrangle_{HQ} = 
\frac{1}{\llangle W_C[0,0]\rrangle } \,\llangle W_C[\delta y_1, \delta y_2] \rrangle
\, .
\stp
%As already mentioned, in thermal field theory, the 
%fields evaluated in the second copy of the real axis can be understood
%as independent (type 2) fields. This also means that we can split the
%source $\delta y(t_c)$ in type 1 and 2 accordingly, and so
%\st
%\delta y_2(t)=\delta y(t-i\beta/2)
%\stp
%And thus, the type 2 source $\delta y_2$ is the variation of the 
%contour in the second copy of the real axis.
The momentum  diffusion coefficient may then be written
\st
  \kappa = \lim_{\omega\rightarrow 0} \int dt\, e^{+i\omega t}\,
  \frac{1}{\llangle W_C[0,0]\rrangle}
  \llangle \frac{\delta^2 W_C[\delta y_1, \, \delta y_2]}
{\delta y_2(t)\; \delta y_1(0)} \rrangle
\, .
\stp

\subsection{$\N=4$ Super Yang Mills}
The previous discussion was performed in a 
SU(N) gauge theory. However, we are interested on studying
the interactions of a heavy W boson in a $\mathcal{N}=4$ $U(N)$
gauge theory broken to $U(N-1) \times U(1)$.
The heavy $W$ boson propagator  is given by the  following
Wilson line \cite{Maldacena:1998im,Drukker:1999zq}
\st
\label{SuperWilson}
U=T_C\exp\left\{-i\int ds \left( A_{\mu} \dot{x}^{\mu} +|\dot{x}|\, \Theta^I X^I
\right) \right\} 
\, ,
\stp
where in addition to the gauge fields the adjoint scalars $X^I$ 
couple to the $W$ boson through a Yukawa type interaction.
Here $\Theta^I$ is the VEV angle.

The extension  to $\mathcal{N}=4$ SYM  amounts to
changing the transporters and Wilson lines in 
accord with Eq. \ref{SuperWilson}. In particular, using 
the non-Abelian Stokes theorem \cite{Bralic:1980ra}
the $\N=4$ analog of \Eq{AlmostWilson} 
\st
W_C[\delta y]=T_C \exp\left\{-i\int_C dt \left(A_{0}+ \Theta^I X^I \right)\right\} 
            T_C \exp\left\{i\int_C dt \, \delta y(t) \, {\tilde{\mathcal E}}_{SYM}(t) \right\} 
\, .
\stp
Here we have defined 
\st
{\mathcal E}_{SYM}=E+D_t\left(\Theta^I \, X^I\right)-D_y\left(\Theta^{I} \, X^{I} \right)
\, ,
\stp
and dressed it with transporters as in Eq.(\ref{dressedF}). The necessity 
of the scalar derivatives is obvious after computing a plaquette with
the links defined in \Eq{SuperWilson}.
In the low frequency limit the time derivatives of the 
scalars can be dropped and the QCD-like field strength is 
modified by gradients of the scalar fields.
\section{The AdS/CFT Correspondence}
\label{fluct}

Fundamental Wilson Loops were first computed by 
Maldacena whose work we recall \cite{Maldacena:1998im}.
On the gauge theory side heavy gauge bosons
are introduced through the Higgs Mechanism. Specifically
the  $U(N)$ gauge group is broken to $U(N-1)\times U(1)$ by giving a large 
expectation value to one of the scalars in the theory. The
resulting $W$ boson transforms in the fundamental representation
of the remaining $U(N-1)$ gauge group.  The 
equation of motion for the $W$ boson  is 
\st
 \left(i\partial_t  - M -  A_0 - \theta^I X^I\right)W  = 0  \, ,
\stp
and naturally leads to  the Wilson loops considered
in the previous section.

On the gravity side this Higgs construction corresponds 
to placing one of the $D3$ branes far from the remaining
$N-1$ $D3$ branes. The propagation
of the heavy $W$ bosons is represented by  semi-classical 
strings which are governed by the Nambu-Goto action
\st
    S = \frac{1}{2\pi \alpha'}\int d\tau d\sigma 
       \sqrt{-\mbox{det} G_{MN}\, \partial_a X^M \partial_b X^N}  \; .
\stp 
In the original Maldacena computation, 
the two parallel Wilson lines 
corresponded to a string world sheet  that circumscribed
the trajectory of a $w(\x)$ and $\bar{w}(\bf y)$ bosons on  the 
boundary.
%(We explicitly indicate the $"1"$ in $w_1$ because these operators
%live on the first branch of the Schwinger-Keyldish contour; At zero
%temperature the $"2"$ branch of the Schwinger-Keyldish contour can
%be ignored.) 

Finite temperature is introduced by 
inserting a black hole into the $AdS$ space.  When 
the black hole is viewed in Kruskal coordinates  there 
are two boundaries which are asymptotically AdS. These 
boundaries correspond to the type  ``1"  and ``2" axis of the 
real time partition function 
\cite{Herzog:2002pc,Ross:2005sc}.  
Naturally the thermal 
description of a broken gauge theory at finite temperature
also has $w$ bosons running along the ``1" and ``2" axis.

\begin{figure}
\begin{center}
\includegraphics[width=10cm]{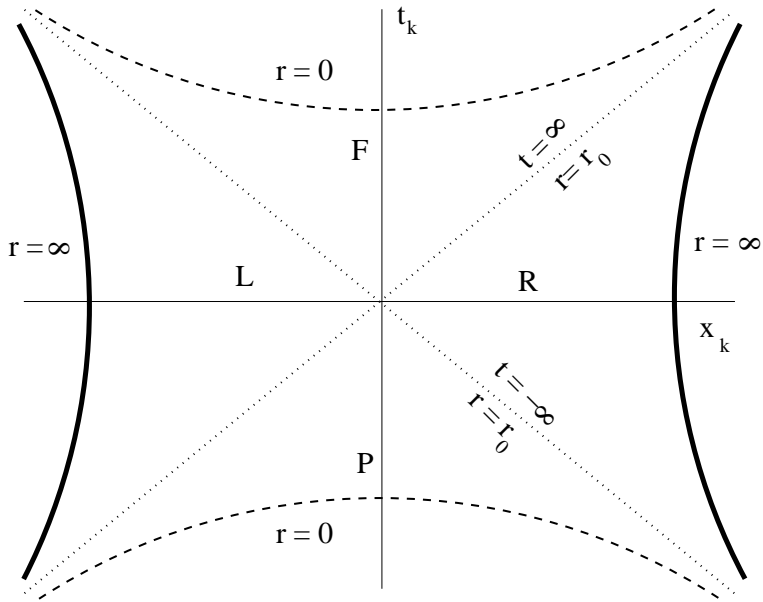}
\caption
{
Kruskal diagram for the AdS black hole. The coordinates $(t,r)$ span the 
right (R) quadrant. The dotted lines and the dashed hyperbolas 
represent the future and past 
horizons and the singularities, respectively. The thick hyperbolas on the sides
of the two quadrants
are the
boundaries  at $r=\infty$. The Wilson line running along the 
Schwinger-Keldysh contour runs along along the ``1" axis (the R boundary) and 
``2" axis (the L boundary).  
This corresponds to a string whose endpoints 
follow these boundaries. The minimal surface with these boundary conditions
is the full Kruskal plane. 
}
\label{kruskal}
\end{center}
\end{figure}

Since we are considering a single heavy gauge boson which propagates along the
full Schwinger-Keldysh contour, there is a single Wilson line on the ``1" axis
and a single Wilson line on the ``2" axis.  It is natural to look for string
solutions in the full Kruskal plane.  Such solutions  connect
the type ``1" $w$ boson at the boundary of the right quadrant with the type
``2" boson at the boundary of the left quadrant -- see \Fig{kruskal}.

%The right (R) and left (L) quadrants of the 
%Kruskal plane correspond to insertion of $1$ type and $2$ 
%type operators into the $SYM$ real time partition function. 
%By placing a single test brane at the boundary, an observer in the
%right (left) quadrant does not see the piece of the boundary brane  
%in the left (right) quadrant. Only the Kruskal observer 
%sees both boundaries.

%In the original Maldacena computation 
%two parallel Wilson lines in quadrant the first quadrant 
%corresponded to a string world sheet  which circumscribes
%the trajectory of the $w_1(\x)$ and $\bar{w}_1(\bf y)$ bosons on  the 
%boundary. 
%(We explicitly indicate the $"1"$ in $w_1$ because these operators
%live on the first branch of the Schwinger-Keyldish contour; At zero
%temperature the $"2"$ branch of the Schwinger-Keyldish contour can
%be ignored.)
%Similarly we must find the string world sheet 
%which circumscribes the the trajectory of the 
%$w_1(\x)$ and $w_2(\x)$ bosons. 

To start, we look for string solutions in the first quadrant.
The metric is 
\st
 ds^2 =  \frac{r^2}{R^2}\left[-f(r) dt^2  + d\x_{\parallel}^2\right] +  \frac{R^2}{f(r) r^2}\,dr^2 + R^2 d\Omega_5^2
\, ,
\stp
where $f(r) = 1 - \left(\frac{r_o}{r}\right)^4$, $R$ is the $AdS$ radius,  and
$r_o$ is related to the Hawking temperature $\pi T R^2 = r_0$. 
We define the scaled units, $\pi T t = \bar{t}$, $\pi T x = \bar{x}$
and $\bar{r} = r/r_o$ and the metric reads
\st
\frac{ds^2}{R^2} = -\bar{r}^2\, f(\bar{r})\, d\bar{t}^2 + \bar{r}^2 
d\bar{\x}_{\parallel}^2 +  \frac{d\bar{r}^2}{f(\bar{r})\,\bar{r}^2} + 
d\Omega_5^2 \, .
\stp
In what follows we will drop the ``bar" until \Eq{Gret}

Parameterizing the string as 
\[
   (\tau,\sigma)\mapsto (t=\tau,\, \x_{\parallel}(\tau,\sigma),\,  r=\sigma,\, \Omega_5(\tau,\sigma) ) \, ,
\]
It is straight forward to show and well known that 
\[
   \x_{\parallel} = \mbox{Const} \qquad \mbox{and} \qquad \Omega_5 =\mbox{Const} \, , 
\]
is a solution to the equations of motion.
This solution extends to the full Kruskal plane and
is the string solution which we are looking for.

To
see this we first write the metric in Kruskal coordinates
\footnote{
Our notation closely follows \Ref{Fidkowski:2003nf}, appendix A. 
Define
$r_{*} \equiv \int_0^{r} \frac{dr}{r^2 f(r)) } + i\frac{\pi}{4}$ so  that $r^{*}$ is real in the right quadrant. 
The Kruskal coordinates are 
$-e^{4r_*}= t_k^2  - x_k^2$ and  $2t = \tanh^{-1}\left(t_k/x_k\right)$
}
\st
   \frac{ds^2}{R^2} = g_k (-dt_k^2 + dx_k^2) + r^2\,d\x_{\parallel}^2 + d\Omega_5^2
\, ,
\stp
where the Kruskal scale factor is $g_k =(1/4)\,r^2 f(r)\, e^{-4r_*} $.
Parametrizing the string as
\st
   (\tau,\sigma) \mapsto (t_k=\tau,\,\x_{\parallel}(\tau,\sigma),\,  x_k=\sigma,\, \Omega_5(\tau,\sigma) ) \, ,
\stp 
we determine the induced metric 
\st
  {\mathcal L} = \sqrt{-\mbox{det} h} = R^2 \left[g^2_k +g_k
                                              \left(r^2 (\x_{\parallel}')^2 - r^2 (\dot{\x}_{\parallel})^2
                                                  +  (\Omega_5')^2  - ({\dot{\Omega}}_5)^2\right)\right]^{1/2}
\, ,
\stp
where, for instance, $\dot{\x}_{\parallel}= \partial_\tau \x_{\parallel}$
and ${\x}'_{\parallel} = \partial_\sigma \x_{\parallel}$.
Then we compute the canonical momenta
\bg
  \frac{\delta \mathcal{L}}{\delta (\dot{\x}_{\parallel})} 
&=& -\frac{ g_k\,r^2 \, \dot{\x}_{\parallel}}{\sqrt{-\mbox{det}h}} 
\, ,
\\
  \frac{\delta \mathcal{L}}{\delta (\x'_{\parallel})} 
&=& \frac{ g_k\,r^2 \, \x'_{\parallel}}{\sqrt{-\mbox{det}h}} 
\, ,
\nd
and we see that the equations of motion
\[
  \partial_{\tau} \frac{\delta \mathcal{L}}{\delta \dot{\x} } 
  + \partial_{\sigma} \frac{\delta \mathcal{L}}{\delta \x_{\parallel}'} =0
\, ,
\]
are satisfied trivially by  $\x_{\parallel} = \mbox{Const}$.  The
canonical momenta vanish.  A similar remark applies to the
$\Omega_5$ coordinates. Thus 
\st
  \x_{\parallel}(t_k,x_k) = \mbox{Const}\,, \qquad \Omega_5(t_k,x_k) = \mbox{Const} \, ,
\stp
is a solution to the string equations of motion throughout the 
Kruskal plane. This represents a Wilson line along the ``$1$" time axis
and  back along the ``$2$" axis of the Schwinger-Keldysh contour.

Having constructed the string solution representing a $W$ 
boson propagating all along the Schwinger-Keldysh contour 
we will proceed to fluctuate the shape of this string solution.
Following the general philosophy of the $AdS/CFT$ correspondence
we equate the exponential of the classical action of the source to the 
generating functional 
\st
\label{generating}
     \frac{1}{e^{iS_{NG}[0,0]}} e^{iS_{NG}[\delta y_1, \delta y_2]}  
=  \frac{1}{\llangle W[0,0] \rrangle} \, \llangle W[\delta y_1, \delta y_2] \rrangle
\, .
\stp
Now that the problem has been formulated along the Schwinger-Keldysh
contour and KMS is satisfied, the results of Son and Herzog \cite{Herzog:2002pc} 
may be adopted $mutatis$ $mutandis$. 
(For related work,  see Ref.~\cite{Ross:2005sc,Unruh:1976db,Israel:1976ur}.) 
For the 
retarded correlator, 
\st
      G_R(\omega) = -i\int_0^{\infty}dt\, e^{+i\omega t}\theta(t)\llangle \left[ \mathcal F(t), \mathcal F(0) \right]\rrangle_{HQ} \, ,
\stp
the procedure is the following. In the 
right Kruskal quadrant, solve the classical equations
of motion for the source $\delta y(t,u)$ with
infalling boundary conditions at the horizon. The retarded correlator 
may be found by evaluating the classical action 
of the string and  differentiating twice with respect to the 
the boundary source, $\delta y(t,u=0)$.

Returning to the first quadrant with  $t,\x_{\parallel},u\equiv \frac{1}{r^2}$  
coordinates, the relevant string action
with small fluctuations  in the y-direction is
\st
%S_{NG} = \frac{R^2}{2\pi \alpha'} 
%\left(\pi T)^3 \, \int \frac{dt\, du}{2u^{3/2}} \, 
%\left[1 - 
%\frac{1}{2} \left( \frac{\dot{\y}_{\parallel}^2}{f} - 
%            f r^4 \left(\partial_u \y_{\parallel'}^2\right) \right)  
% - \frac{1}{2} \left(\frac{\dot{\Omega_5}^2 u}{f} - 
%              4 f u^2 (\partial_u\Omega_5)^2\right) 
%\right]
S_{NG}=\frac{R^2}{2\pi\alpha'}
\int \frac{dt\,d u}{2u^{3/2}}
\left[
1-\frac{1}{2}\left(
             \frac{\dot{y}_{\parallel}^2}{f}-
              4 f u \left(y'_{\parallel}\right)^2
             \right)
\right]
\, .
\stp

Notice that the infinite action corresponding to the
unperturbed string is to be subtracted since it 
appears in the numerator and denominator of \Eq{generating}.
Fluctuating in the $y$ direction, we define
\st
  y(t,u) = \int e^{-i \w t}\, y(\w)\, Y_\w(u)  \, \frac{d\w}{2\pi}
\, ,
\stp
and the  equation of motion of the fluctuating string is
\st
\partial_{u}^2 Y_\w - \frac{(2 + 6 u^2)}{4 u f} \partial_u Y_\w + 
\frac{\w^2}{4 u f^2 } Y_\w = 0
\, .
\stp

This equation has  regular singular points at 
the boundary $u=0$ and at the horizon $u=1$. 
Near the horizon the solution behaves as 
\st
%      \delta
 Y_\w(u) = (1-u)^{\pm i \w/4} .
\stp
Remembering that $u\equiv 1/r^2$, these
solutions correspond to outgoing and infalling fluctuations.
For the 
retarded propagator, 
\st
      G_R(\omega) = -i\int_0^{\infty}dt\, e^{+i\omega t}\theta(t)\llangle \left[ \mathcal F(t), \mathcal F(0) \right]\rrangle_{HQ} \, ,
\stp
incoming boundary conditions must be selected for the 
corresponding source, 
%$\delta z_{\omega}(u)$.  
$Y_{\w}(u)$.  
We therefore substitute 
$Y_{\w}(u) = (1 -u^2)^{-i\w/4}\,F_{\w}(u)$
where $F_{\w}(u)$ is regular at the horizon. Since $F_{\w}(u)$ 
is regular at the horizon we may expand in a power series 
in $\w$ and solve order by order
\footnote
{
We thank A. Starinets for providing us this expression.
}
\begin{eqnarray}
\label{solution}
    Y_\w(u) &=& (1-u^2)^{-i\frac{\w}{4}} \times \\
                & & \left[ 1 + 
\frac{i \w}{8} \left\{\pi - 4\tan^{-1}(\sqrt{u}) - 6\log 2 + 4\log(1+\sqrt{u}) + 2\log(1+u) \right\}  + 
 O(\w^2)  \right]
\, .
\nonumber
\end{eqnarray}
The insertion of this solution into the action 
 reduces to a boundary term. 
After re-inserting units 
\[
y_\omega (\omega)=(\pi T)^2 \bar{y}_{\w} (\w=\frac{\omega}{\pi T}) \qquad Y_{\omega}(\omega,u) = \bar{Y}_{\w}(\w =\frac{\omega}{\pi T},u) \, ,
\]
and denoting $A(u)$ as the coefficient in front of the 
kinetic term as in Refs.~\cite{Policastro:2002se,Herzog:2002pc} 
we have 
\st
\label{Gret}
     G_{R}(\omega) = - A(u) Y_{-\omega}(u) \partial_u Y_\omega(u)|_{u\rightarrow 0} \, ,
  \qquad   A(u) = \frac{R^2\, \left(\pi T \right)^2}{\pi \alpha'}
                  \frac{f}{u^{1/2}}
\, .
\stp
Substituting the solution \Eq{solution} into this expression 
we obtain
\begin{eqnarray}
\label{kappa}
  \kappa  &=& \lim_{\omega\rightarrow 0} \,\frac{-2T}{\omega}\, \mbox{Im} G_{R}(\omega)\,, \\
          &=& \sqrt{ \lambda}\, T^3\,\pi  \, .
\end{eqnarray}
The divergence which arises as $u\rightarrow 0$ does not
effect the imaginary part. In the last step we have used
the relation $R^2/\alpha' = \sqrt{\lambda}$.
The significance of the diffusion  computation  will be discussed in the next section.

\section{Discussion}

\label{discuss}
In summary we have computed the heavy quark diffusion coefficient
in $\N=4$ SYM by exploiting the $AdS/CFT$ correspondence.
From \Eq{kappa} and \Eq{DfromKappa}  the 
diffusion coefficient is
\begin{eqnarray}
\label{diffusion}
    D &=& \frac{2 T^2}{\kappa} \, \\
      &=& \frac{2}{\pi T} \frac{1}{\sqrt{\lambda}}
\, ,
\end{eqnarray}
where $\lambda = g^2_{YM} N_c$. 
This  result is interesting both  theoretically and phenomenologically.

An immediate observation is that the heavy quark diffusion
coefficient $D$  is parametrically small compared to the momentum diffusion coefficient
$\eta/(e+p)=1/(4\pi T)$ \cite{Policastro:2001yc} , 
{\it i.e.} $D$ depends on the peculiar, but 
characteristic, $1/\sqrt{\lambda}$.  This should be contrasted 
with perturbation theory where
all transport scales are the same order of magnitude,  
$1/(\lambda^2 T)\,\log(\lambda^{-1})$.
%At strong coupling, the large  separation between the 
%electric field relaxation rate
%$\sqrt{\lambda} T$, and the 
%momentum relaxation rate $T$ 
%The differences between these suggests a strong coupling approximation. 
%Perhaps the difference of time scales can be used to set up an
%effective theory of $\N=4$ with a few coefficients calculated with
%gravity duals.

Theoretically, the diffusion computation proceeded as follows.  A Wilson line running 
along the Schwinger-Keldysh contour is the partition function of a single heavy quark 
\footnote{ The shape of the contour is irrelevant.  Taking a contour which
runs straight down  the imaginary time axis to $-i\beta$ gives the 
Polyakov loop.  This is the more common but equivalent definition of the heavy quark partition function. }.
There is a Wilson line on the ``1" axis and a Wilson line on the ``2" axis.
In a
Minkowski metric  the corresponding semi-classical string solution must span the
full Kruskal plane if the right (left) quadrant of the Kruskal diagram is to be
identified with  ``1" (``2") fields in the real time path integral. We find this
semi-classical solution in \Sect{fluct},  and it is   a single straight string in $AdS_5 \times
S_5$ which extends from the boundary to the event horizon when viewed from an
observer in the right quadrant.  Force-force correlators in the presence of a
heavy quark are electric field correlators with links running  along the
contour after the heavy quark fields have been integrated out.  These same
correlators may be obtained by variations of the unperturbed Wilson loop. Thus
by varying the end point  of the relevant string solution in $AdS\times S_5$
one may compute real time force-force correlators, $i.e.$ the diffusion
coefficient.

Given the diffusion coefficient in \Eq{diffusion}, 
the quark must be sufficiently heavy for the Langevin theory to 
apply. Since the Langevin theory is classical, consistency demands 
that the relaxation time be large compared to the inverse
temperature  
\st
       \frac{M}{T}\,D \gg \frac{1}{T} 
\, .
\stp
This leads  to the following constraint
\st
       M \gg \frac{\pi T}{2} \, \sqrt{\lambda}
\, .
\stp
For a string of length $L$ stretching 
outward from the horizon we have,
$M\sim L/\alpha'$, $\pi T = r_o/R^2_{\scriptscriptstyle AdS}$, and $\sqrt{\lambda}=R^2_{\scriptscriptstyle AdS}/\alpha'$.   This Langevin constraint is satisfied
by the gravity computation whenever
\st
     L \gg r_o  \, .
\stp
$i.e.$ whenever the length of the string is long enough.  
Clearly mass or energy of order $T\sqrt{\lambda}$ sets a 
boundary for a quasi-particle picture to be valid. In the
gravity computation this corresponds to a string of length $L\sim r_o$.

Phenomenologically the diffusion coefficient is an extremely 
interesting number. Bearing in mind that $\N=4$ SYM is not 
QCD, the authors still want to substitute numbers
\st
      D \simeq \frac{0.9}{2\pi T} \, \left(\frac{1.5}{\alpha_{SYM} N} \right)^{1/2}  
\, .
\stp
This could be compared to extrapolations 
of weak coupling to QCD \cite{Moore:2004tg,Braaten:1991we} 
\st
     D \simeq \frac{6}{2\pi T}\, \left(\frac{1.5}{\alpha_s N_c}\right)^2
\, .
\stp
Of equal importance is the mass scale where we expect the heavy quark
theory to apply
\st
     M  \gg   1.7 \mbox{GeV}\, \left( \frac{T}{0.250\, \mbox{GeV}} \right)\, \left( \frac{ \alpha_{SYM} N }{1.5} \right)^{1/2} \, .
\stp
This suggests that the Langevin process might not be 
applicable to charm quarks, though of course there is  an
unknown proportionality factor in this equation.

The RHIC data favor a diffusion coefficient of about $3/(2\pi T)$.
However, relativistic effects obscure the relation between 
the diffusion coefficient and the semi-leptonic data.  
The role of radiative and collisional energy loss remains 
unclear \cite{Djordjevic:2005db,Armesto:2005zy}. 
The next step from the perspective of 
the $AdS/CFT$ correspondence is to consider quarks with 
finite velocity -- work is in progress in this 
direction \cite{get-crackin}. The framework set up here should 
allow a computation of the real time transport properties of 
these quarks.

\noindent {\bf Note Added.}
In the  few days surrounding this preprint
three similar studies of heavy quark energy loss appeared.
The first of these computed the transverse momentum 
diffusion of an ultra-relativistic quark \cite{Wiedemann}.
The second paper, denoted HKKKY after the authors, 
computed the drag on a quark moving 
with finite velocity \cite{HKKKY}. A third 
paper \cite{Gubser:2006bz} independently computed the
drag using methods similar to HKKKY and obtained the 
same answer. The setup in these papers differs
signficantly from this work.

The HKKKY paper
is particularly instructive.  Consider the Langevin 
process of a heavy quark 
\begin{eqnarray}
\frac{dp_i}{dt} &=& \xi_i(t) - \eta_D p_i \; , \qquad \langle \xi_i(t) \xi_j(t') \rangle = \kappa \delta_{ij}
	\delta(t-t') \, ,
\end{eqnarray}
where the drag,  fluctuations, and diffusion are related by
Einstein relations
\[
   \eta_{D} = \frac{\kappa}{2MT}\,, \qquad D = \frac{2T^2}{\kappa} \, .
\]
In this paper we considered times short compared
to the relaxation time $(M/T)\,D$, and computed the 
strength of the noise $\kappa$, taking the mass to infinity.
The mass is not needed to compute the diffusion coefficient or
$dE/dx = T/D$.

HKKKY computed the drag by considering the change in 
the average momentum over a time long compared to the 
time scale of noise correlations, $\sim 1/T$.  The Langevin process (see e.g. \cite{Moore:2004tg}) predicts  
the probability that a quark with momentum $\p_0$  at
time $0$ will arrive with momentum $\p$ at later time $t$
\st
\label{Green}
P(\p, t | \p_0,  0)
 =
 \frac{ 1}{\sqrt[3]{ 2 \pi M T\, (1 - e^{-2 \eta_D t}) }  }
 \exp\left[ - \frac{ (\p - \p_0 e^{-\eta_D t})^2 }{2 M T\, (1-e^{-2 \eta_D t})} \right] \, . \nonumber \\
\stp
The average momentum obeys
\st
    \llangle \p(t) \rrangle = \p_{0}e^{-\eta_D t} \, , 
\stp
while the width obeys
\st
   \llangle (\Delta \p)^2 \rrangle = 3 M T\, \left(1 - e^{-2\eta_D t} \right) \, .
\stp
HKKKY considered an atypical quark, with momentum $p_0 \gg \sqrt{MT}$,
and computed the change  in the average momentum over 
a time interval of order, $1/\eta_D = (M/T)\,D$. This determines the drag, $\eta_D$. For an {\it atypical} quark  
the fluctuations are neglible over this time interval.
They also compute the mass and
 then are  able to deduce diffusion coefficient through the Einstein
relations.
The result agrees with \Eq{diffusion};   the $AdS/CFT$ correspondence is
neatly consistent with the fluctuation dissipation theorem. 
It 
would be quite interesting to see the full structure of the 
Langevin Green function emerge from the string 
theory. In essence this computation has been performed already
through an amalgamation of our works.

HKKKY also computed the heavy quark energy loss at finite 
velocity. The bending string solution they obtained 
agrees with our preliminary computations.

\noindent {\bf Acknowledgments.} We wish 
to thank Andrei Starinets and Edward Shuryak for 
fruitful discussions.
JC thanks M. Kulaxizi, S. Giombi, R. Ricci and D. Trancanelli.
We especially thank
Ismail Zahed for challenging us to find a clean formulation
of the problem. This work   
was supported by grants from the U.S. Department of Energy,
DE-FG02-88ER40388 and DE-FG03-97ER4014.  

\appendix
\section{The Heavy Quark Partition Function}
\label{partition}
In this section we clarify several aspects concerning the
heavy quark partition function. We follow closely the 
discussion of McLerran and Svetitsky \cite{McLerran:1981pb}. We introduce the operators $Q({\bf x},t)$,
$Q^{\dagger}({\bf x},t)$ which create and annihilate static quarks at point
$\x$ and time $t$. These fields satisfy the anti-commutation relations
\st
\{Q_i(\x,t),Q_j^{\dagger}(\x',t)\}=\delta_{ij}\delta(\x-\x') \, .
\stp
From the Lagrangian \Eq{lagrangian} the time evolution of
quark fields is given by 
%\st
% \left(i\partial_{t_C}  -M -  A_0 \right)Q(\x,t)  = 0 
%\stp
%which can be easilly integrated to 
\st
\label{integrated}
Q(\x,t)=e^{-i\,M (t-t_0)}U(t,t_0) Q(\x,t_0) \, ,
\stp
where both $t$ and $t_0$ are in the Schwinger-Keldysh contour
and $U(t,t_0)$ is the transporter between these two points
at fixed position $\x$. As in Ref. \cite{McLerran:1981pb}, the
partition of function in the presence of a heavy quark is
\begin{eqnarray}
Z_{HQ}=\sum_s \llangle s \left| e^{-\beta H} \right | s \rrangle &=&
    \int d^3 x   \sum_{s'} \llangle  s' \left | Q(\x, -T) e^{-\beta H}  
                                              Q^{\dagger}(\x,-T) 
                  \right | s' \rrangle \, ,  \\
    & = &
    \int d^3 x   \sum_{s'} \llangle  s' \left | e^{-\beta H} Q(\x,-T-i\beta) 
                                              Q^{\dagger}(\x,-T) 
                  \right | s' \rrangle 
\, ,
\end{eqnarray}
in which $\left|  s \right>$ is a state of the system with only 
one 
heavy quark, and $\left|  s' \right>$  is a state with no heavy quarks 
({\it i.e.} $Q \left|  s' \right>=0$). By means of Eq. 
(\ref{integrated}) we obtain Eq. (\ref{ZHQ}).
With this definition of the partition function, the force-force correlator is
\st
\label{Dg}
D^{>}(t)=\llangle \mathcal{F}(t) \mathcal{F}(0) \rrangle _{HQ}=
\frac{1}{Z_{HQ}}
\sum_s \llangle s \left| e^{-\beta H} \mathcal{F}(t) \mathcal{F}(0)
\right | s \rrangle
\, .
\stp
We introduce a complete sets of states between the 
two force operators. We notice that, since we are assuming 
weak fields (which cannot create heavy quarks) and 
since the force operator does not change the number
of quarks, it is enough to consider a complete set of one particle heavy 
quark states plus bath. Following  Ref.
\cite{LeBellac} we obtain the KMS relation,
\st
D^{>}(t)=\llangle \mathcal{F}(0) \mathcal{F}(t+i\beta) \rrangle _{HQ}
        =D^{<}(t+i\beta)
\, .
\stp
Since the force is a Hermitian operator  local in time, the definition of the
spectral density $\rho(t)=D^{>}(t)-D^{<}(t)$ leads to the standard
relations between correlators, \Eq{G11}, (\ref{G12}) and (\ref{G22}).

We conclude by remarking that, since the states $\left|  s' \right>$
can be considered as the $\left|  0_{A} \right>$ Fock states for the
heavy quarks in a gauge field background, the partition function is
\st
\label{vacPI}
Z_{HQ}=\int d^3 x \Tr \left[ \left<  0_{A} \right| Q(\x,-T-i\beta)   Q^{\dagger}(\x,-T)
                \left|  0_{A} \right> \right] 
\, ,
\stp
where the $\Tr$ represents the thermal trace over the bath (the Yang-Mills
fields). Thus, representing the vacuum expectation value in 
\Eq{vacPI} as a path integral we arrive at expression 
\Eq{ZHQPI}. The effect of the vacuum average is to 
introduce  the $i\epsilon$ prescription in Eq. (\ref{GF}) as in
zero temperature field theory \cite{Weinberg}.

\end{document}